\documentclass[11pt]{article}
\usepackage{epsfig}
\usepackage{graphicx}
\usepackage{latexsym}
\usepackage{amssymb}
\RequirePackage{mathptmx}
\RequirePackage[T1]{fontenc}
\usepackage{heppennames2}
\usepackage{lhchiggs}
\usepackage{cernunits}
\usepackage{ctable}
\usepackage[center,small,bf]{caption}
\DeclareSymbolFont{forjmath}{OT1}{cmr}{m}{sl}
\DeclareMathSymbol{\Jmath}{\mathord}{forjmath}{'021}
\def\jmath{\Jmath}
\DeclareFontFamily{OT1}{cmr}{}
\DeclareFontFamily{OT1}{cmss}{}
\allowdisplaybreaks
\newcommand\pubdate{\today}

\def\csumb{Dipartimento di Fisica Teorica, Universit\`a di Torino, Italy\\
           INFN, Sezione di Torino, Italy}
\def\support{\footnote{Work supported by MIUR under contract
    2001023713$\_$006 and by Compagnia di San Paolo under contract ORTO11TPXK.}}
\def\Title#1{\begin{center} {\Large\bf #1 } \end{center}}

\def\Author#1{\begin{center}{ \sc #1} \end{center}}
\def\Address#1{\begin{center}{ 
\normalsize \bfseries \itshape #1} \end{center}}

\newcommand\pubblock{\rightline{\begin{tabular}{l} \\
         \pubdate\\  \end{tabular}}}
\newenvironment{Abstract}{\begin{quotation}  }{\end{quotation}}

\def\email#1{\footnote{\tt EMAIL: #1}}
\makeatletter
 \def\citenum#1{{\def\@cite##1##2{##1}\cite{#1}}}
\def\citea#1{\@cite{#1}{}}
\newcount\@tempcntc
\def\@citex[#1]#2{\if@filesw\immediate\write\@auxout{\string\citation{#2}}\fi
  \@tempcnta\z@\@tempcntb\m@ne\def\@citea{}\@cite{\@for\@citeb:=#2\do
    {\@ifundefined
       {b@\@citeb}{\@citeo\@tempcntb\m@ne\@citea\def\@citea{,}{\bf }\@warning
       {Citation `\@citeb' on page \thepage \space undefined}}%
    {\setbox\z@\hbox{\global\@tempcntc0\csname b@\@citeb\endcsname\relax}%
     \ifnum\@tempcntc=\z@ \@citeo\@tempcntb\m@ne
       \@citea\def\@citea{,}\hbox{\csname b@\@citeb\endcsname}%
     \else
      \advance\@tempcntb\@ne
      \ifnum\@tempcntb=\@tempcntc
      \else\advance\@tempcntb\m@ne\@citeo
      \@tempcnta\@tempcntc\@tempcntb\@tempcntc\fi\fi}}\@citeo}{#1}}
\def\@citeo{\ifnum\@tempcnta>\@tempcntb\else\@citea\def\@citea{,}%
  \ifnum\@tempcnta=\@tempcntb\the\@tempcnta\else
  {\advance\@tempcnta\@ne\ifnum\@tempcnta=\@tempcntb \else\def\@citea{--}\fi
    \advance\@tempcnta\m@ne\the\@tempcnta\@citea\the\@tempcntb}\fi\fi}
\newcommand\KW[1]{\vskip .5pt \vbox{\small Keywords:\ #1}}
\newcommand\PACS[1]{\vskip .5pt \vbox{\small PACS:\ #1}}
\newcommand\secstyle{\bfseries\raggedright}
\renewcommand\section{\@startsection{section}{1}{\z@}%
                                   {3.5ex \@plus 1.3ex \@minus .7ex}%
                                   {2.3ex \@plus.4ex \@minus .4ex}%
                                   {\normalfont\large\secstyle}}
\renewcommand\subsection{\@startsection{subsection}{2}{\z@}%
                                   {2.3ex\@plus 1ex \@minus .5ex}%
                                   {1.2ex \@plus .3ex \@minus .3ex}%
                                   {\normalfont\normalsize\secstyle}}
\renewcommand\subsubsection{\@startsection{subsubsection}{3}{\z@}%
                                   {2.3ex\@plus 1ex \@minus .5ex}%
                                   {1ex \@plus .2ex \@minus .2ex}%
                                   {\normalfont\normalsize\secstyle}}%
\makeatother
\DeclareRobustCommand{\PA}{\HepParticle{A}{}{}\Xspace}

\DeclareRobustCommand{\PV}{\HepParticle{V}{}{}\Xspace}

\DeclareRobustCommand{\PX}{\HepParticle{X}{}{}\Xspace}

\DeclareRobustCommand{\Pf}{\HepParticle{f}{}{}\Xspace}
\DeclareRobustCommand{\PAf}{\HepAntiParticle{\Pf}{}{}\Xspace}
\DeclareRobustCommand{\PF}{\HepParticle{F}{}{}\Xspace}

\DeclareRobustCommand{\Ph}{\HepParticle{h}{}{}\Xspace}

\newcommand{\myLO}{\rm{\scriptscriptstyle{LO}}}
\newcommand{\myNLO}{\rm{\scriptscriptstyle{NLO}}}

\newcommand{\myEW}{{\mathrm{EW}}}
\newcommand{\myQCD}{{\mathrm{QCD}}}

\newcommand{\ssF}{{\mathrm{F}}}
\newcommand{\ssR}{{\mathrm{R}}}

\newcommand{\ssS}{{\mathrm{S}}}

\newcommand{\bqas}{\begin{eqnarray*}}
\newcommand{\eqas}{\end{eqnarray*}}

\newcommand{\lpar}{\left(\Xspace}                            
\newcommand{\rpar}{\Xspace\right)}

\newcommand{\bq}{\begin{equation}}                    
\newcommand{\eq}{\end{equation}}
\newcommand{\bqa}{\arraycolsep 0.14em\begin{eqnarray}}
\newcommand{\eqa}{\end{eqnarray}}
\newcommand{\ba}[1]{\begin{array}{#1}}
\newcommand{\ea}{\end{array}}
\newcommand{\ben}{\begin{enumerate}}
\newcommand{\een}{\end{enumerate}}
\newcommand{\bei}{\begin{itemize}}
\newcommand{\eei}{\end{itemize}}
\newcommand{\eqn}[1]{Eq.(\ref{#1})}

\newcommand{\ord}[1]{{\mathcal O}\lpar#1\rpar}

\newcommand{\Bref}[1]{Ref.~\cite{#1}}
\newcommand{\Brefs}[1]{Refs.~\cite{#1}}

\newcommand{\eg}{e.g.\xspace}
\newcommand{\ie}{i.e.\xspace}
\newcommand{\etc}{etc.\@\xspace}

\newcommand{\mh}{\mathswitch {M_{\PH}}}
\newcommand{\mw}{\mathswitch {M_{\PW}}}

\newcommand{\mz}{\mathswitch {M_{\PZ}}}
\newcommand{\mpf}{\mathswitch {M_{\Pf}}}
\newcommand{\tgw}{\mathswitch {\Gamma_{\PW}}}
\newcommand{\tgz}{\mathswitch {\Gamma_{\PZ}}}

\newcommand{\mt}{\mathswitch {M_{\PQt}}}
\newcommand{\mb}{\mathswitch {M_{\PQb}}}

\newcommand{\shat}{\mathswitch {\hat s}}
\newcommand{\muR}{\mathswitch {\mu_{\ssR}}}
\newcommand{\muF}{\mathswitch {\mu_{\ssF}}}

\newcommand{\tot}{{\mbox{\scriptsize tot}}}

\DeclareRobustCommand{\PK}{\HepParticle{K}{}{}\Xspace}

\DeclareRobustCommand{\PAf}{\HepAntiParticle{\Pf}{}{}\Xspace}

\DeclareRobustCommand{\Ph}{\HepParticle{h}{}{}\Xspace}

\begin{document}
\begin{titlepage}
\pubblock
%
\vfill
\def\thefootnote{\fnsymbol{footnote}}
\Title{\LARGE \sffamily \bfseries
Higgs Boson Production and Decay:\\[0.2cm]
Dalitz Sector\support
}
\vfill
\Author{\normalsize \bfseries \sffamily
Giampiero Passarino \email{giampiero@to.infn.it}}               
\Address{\csumb}
\vfill
\vfill
\begin{Abstract}
\noindent 
The processes $\PH \to \PAf\Pf\PGg(\Pg)$, $\Pp\Pp \to \PAQq + \PQq \to \PH +\PGg(\Pg)$ and
$\Pp\Pp \to \PQq(\PAQq) + \Pg \to \PH + \PQq(\PAQq)$ pose severe challenges to the experimental 
analysis. They represent rare decays and production mechanisms of the Higgs boson at LHC. 
However, they are not Yukawa suppressed at next-to-leading order opening a window for the correct 
definition of pseudo-observables, \eg a definition of $\Gamma\lpar \PH \to \PZ\PGg\rpar$
with universal inherent meaning, that are currently used in extracting information for the 
couplings of the newly discovered resonance at LHC. The impact of genuinely electroweak NLO 
corrections is discussed, as well as the comparison of
$\sigma\lpar \Pp\Pp \to \Pg\Pg\PX \to \Pep\Pem\PGg\rpar$ to its zero-width approximation.
\end{Abstract}
\vfill
\begin{center}
\KW{Feynman diagrams, loop calculations, radiative corrections, Higgs physics} 
\PACS{12.15.Lk 11.15.Bt, 12.38.Bx, 13.85.Lg, 14.80.Bn, 14.80.Cp}
\end{center}
\end{titlepage}
\def\thefootnote{\arabic{footnote}}
\setcounter{footnote}{0}
\small
\thispagestyle{empty}
%
\normalsize
\clearpage
\setcounter{page}{1}
\section{Introduction}
The original motivation for this paper is on interpretation of the pseudo-observable (PO)
$\PH \to \PZ \PGg$ which is one of the key ingredients, with $\PH \to \PGg\PGg$, in 
studying Higgs boson couplings at LHC, see 
\Brefs{LHCHiggsCrossSectionWorkingGroup:2012nn,Heinemeyer:2013tqa,Passarino:2012cb}. 
For recent and past developments on the experimental side we quote 
\Brefs{ATLAS:2013rma,Chatrchyan:2013vaa,Abazov:2008wg}.

The $\PZ$ boson is an unstable particle, predominantly decaying into a $\PAf\Pf\,$-pair, so that
the Higgs Dalitz decay, $\PH \to \PAf\Pf\PGg$, is the process to be compared with the data;
original work along these lines can be found in 
\Brefs{Abbasabadi:1996ze,Abbasabadi:2006dd,Abbasabadi:2004wq,Dicus:2013ycd}
(see also \Bref{Chen:2012ju}).
There are important points to keep in mind when discussing Dalitz decay of the Higgs boson, in 
particular that the next-to-leading (NLO) electroweak (EW) and QCD corrections are not Yukawa
suppressed~\cite{Passarino:1987za}, contrary to what happens in lowest order (LO). Therefore, 
we have extended the analysis to cover all related processes that share this property, 
$\PH \to \PAQq\PQq\Pg$ and Higgs - photon(gluon) associated production at hadron colliders. For 
the original work on NLO EW corrections to Higgs - gluon associated production we quote 
\Bref{Keung:2009bs} (see also \Bref{Anastasiou:2011pi}). For the inclusion of $\PQb$ quarks, see
\Bref{Brein:2010xj}.

Returning to the original question of how to link the pseudo-observable $\Gamma(\PH \to \PZ\PGg)$ 
to a specific set of experimental data, we observe the following: it came dangerously close to 
realizing a nightmare, physics done by sub-sets of diagrams (\eg $\PH \to \PZ\PGg$) instead of 
kinematical cuts (\eg on $\PH \to \PAf\Pf\PGg$). Several years ago we avoided that 
fate~\cite{Bardin:1999gt,Bardin:1999ak}, may be the history will repeat itself?

Why Dalitz decay? For a Standard Model (SM) Higgs boson of $125.5\UGeV$ we find
$\mbox{BR}\lpar \PH \to \Pep\Pem\rpar = 5.1\,\times\,10^{-9}$,
while a naive estimate gives
$\mbox{BR}\lpar \PH \to \PZ\PGg\rpar\,\times\,\mbox{BR}\lpar \PZ \to \Pep\Pem\rpar =
5.31\,\times\,10^{-5}$, which is $4$ orders of magnitude larger. However, how much of this 
number will be reflected into the corresponding PO, consistently extracted from full Dalitz decay?
Once again, a fully inclusive estimate is given by $\Gamma\lpar \PH \to \Pep\Pem\PGg\rpar = 5.7\%\,
\Gamma\lpar\PH \to \PGg\PGg\rpar$~\cite{Sun:2013rqa} but the question cannot be answered before 
discussing photon isolation\footnote{LHCHXSWG BR Subgroup Meeting: focus on Dalitz decay,
https://indico.cern.ch/conferenceDisplay.py?confId=250520.}.
In the following, we introduce categories: the name ``Dalitz decay'' must be reserved for the 
full process $\PH \to \PAf \Pf \PGg$ and subcategories are defined by:
\[
\begin{array}{ll}
\cmidrule{1-2}
\PH \to \PZ^*\lpar \to \PAf\Pf\rpar + \PGg       & \mbox{unphysical} \\
\PH \to \PGg^*\lpar \to \PAf\Pf\rpar + \PGg      & \mbox{unphysical} \\
\PH \to \PZ_{\rm c}\lpar \to \PAf\Pf\rpar + \PGg & \mbox{PO}         \\
\cmidrule{1-2}
\end{array}
\]
where $\PZ^*$ is the off-shell $\PZ$ boson and $\PZ_{\mathrm c}$ is the $\PZ$ boson
at its complex pole. More generally, for a given massive particle, we define its ``Dalitz sector''
as the one containing all four-body processes involving the particle, a massless gauge boson and 
two massless fermions. Understanding the problem of POs means understanding the difference
between $\PH \to \PAf\Pf$ and $\PH \to \PAf \Pf + n\,\PGg\,$; this is most easily done using
an argument based on the cuts of the three-loop $\PH$ self-energy: only the sum over all cuts 
is infrared and collinear finite so that we must isolate photons, otherwise we will be mixing
different processes, $\PH \to \PAf \Pf$ at next-to-next-to-leading order (NNLO) and
$\PH \to \PAf \Pf \PGg$ at NLO.

One should not get trapped by intuition when dealing with data: the infrared/collinear component 
of the decay will not survive in the limit $\mpf \to 0$ while there are genuinely non-radiative
(QED and QCD) terms surviving the zero-Yukawa limit.
Therefore, only the Dalitz decay has a meaning and it can be differentiated through kinematical
cuts; the most important one is the definition of ``visible photons'' to distinguish between 
different final states, $\PAf\Pf$ and $\PAf\Pf\PGg$. Other cuts can be applied on the
invariant mass $M_{\PAf\Pf}$ to isolate pseudo-observables and one has to distinguish:
a) $\PH \to \PAf\Pf +$ soft(collinear) photon(s) which is part of the real 
corrections to be added to the virtual ones in order to obtain $\PH \to \PAf\Pf$ at (N)NLO;
b) a visible photon and a soft $\PAf\Pf\,$-pair where one probes the Coulomb pole and get
large (logarithmic) corrections that should be exponentiated.

Once again, $\PH \to \PZ^* \PGg \to \PAf\Pf \PGg$ and $\PH \to \PGg^* \PGg \to \PAf\Pf \PGg$ are
unphysical: none of these contributions exists by itself, each of them is not even gauge 
invariant. However, one can put kinematical cuts:
with a small window around the $\PZ\,$-peak the pseudo-observable $\PH \to \PZ_{\rm c} \PGg$
can be enhanced (but there is a contamination due to many non-resonant backgrounds).
One should also beware of generic statements about box contamination in $\PH \to \PZ \PGg$ 
being known to be small and of ad-hoc definitions of gauge-invariant splittings.
Of course, at small di-lepton invariant masses $\PGg^*$ dominates.

Our summary is as follows: $\PH \to \PAf\Pf$ is well defined and $\PH \to \PAf\Pf + \PGg$ ($\PGg$ 
soft+collinear) is part of the corresponding NLO corrections while  $\PH \to \PZ\PGg$ is  
ill-defined, being a gauge-variant part of $\PH \to \PAf\Pf + \PGg$ ($\PGg$ visible) and can 
be extracted (in a PO framework) by imposing cuts on the di-lepton invariant mass.

The outline of the paper is as follows: in \refS{CS} we describe the salient features of the
calculation, in \refS{Dec} we present results for the Higgs boson decay while the
associated production is discussed in \refS{Pro}. Theoretical uncertainties are discussed in
\refS{THU}.
\section{Computational setup \label{CS}}
We compute helicity amplitudes for $\PH + \PAf + \Pf + \PGg(\Pg) \to 0$ 
according to \Brefs{Passarino:1983bg,Passarino:1983zs} and express them in terms of Mandelstam 
invariants.

Loop integrals are treated with a) standard reduction to scalar integrals to be
evaluated analytically, b) BST functional 
relations~\cite{Passarino:2001wv,Ferroglia:2002mz,Actis:2008ts} and numerical evaluation.
Comparison of the two approaches provides a powerful check on the results.
Furthermore, for the EW NLO corrections we use the Complex-Pole scheme (CPS) of 
\Brefs{Actis:2006rc,Actis:2008uh,Passarino:2010qk,Goria:2011wa}; as input parameters for the 
numerical evaluation we have used the following values:
\[
\begin{array}{llll}
\cmidrule{1-4}
\mw = 80.398\,\UGeV  \; & \; 
\mz = 91.1876\,\UGeV \; & \;
\mt = 172.5\,\UGeV  \; & \;
\tgw = 2.0887\,\UGeV\\
G_{\ssF} = 1.16637\,\times\,10^{-5}\,\UGeV^{-2}  \; & \; 
\alpha(0) = 1/137.0359911                   \; & \; 
\alpha_{\ssS}\lpar \mz\rpar=  0.12018    \; & \; 
\tgz = 2.4952\,\UGeV \\
\cmidrule{1-4}
\end{array}
\]
For the PDF we use MSTW2008 at NLO~\cite{Martin:2009iq}. At LO we use $\mb = 4.69\UGeV$
and derive $\Gamma_{\PQt} = 1.480\UGeV$.

Once the helicity amplitudes are computed we use an optimization scheme based on the notion of
``abbreviations''. We are dealing with multivariate polynomials in the Mandelstam variables; we
require their evaluation to be performed with the least number of arithmetic operations and each
polynomial will receive a name (abbreviation). The strategy, also known as ``subexpression 
elimination'', represents a code transformation in which variables are introduced for each
subexpression such that it is calculated only once and can be used at any later point in the
calculation. 

Schematically, invariants are collected (bracketed) and brackets are factored out; the procedure 
is repeated until the innermost brackets contain only monomials or polynomials that are 
irreducible over $R$. The innermost brackets are ``abbreviated'', the next level of brackets 
is again ``abbreviated'' etc. All abbreviations are then pre-computed (once and only once) in 
the numerical code. For an alternative approach we refer to the work in \Bref{Reiter:2009ts};
we mention that for multivariate polynomials there is no a priori knowledge of the scheme that
leads to the smallest number of operations.

Another improvement in calculation speed is given by the introduction of collinear-free
functions~\cite{Bardin:2009zz}. As it is well known~\cite{Dittmaier:2003bc}, infrared/collinear 
singular configurations in one-loop $n\,$-point functions arise only from three-point 
sub-diagrams. The best way of introducing collinear-free functions is given by the BST 
decomposition~\cite{Passarino:2001wv,Ferroglia:2002mz,Actis:2008ts}; for instance, a box diagram 
in four dimensions can be written as a linear combination of a box in six dimensions (which is 
never soft/collinear divergent) plus vertices in four. In this way it is very simple to check 
(analytically) for the cancellation of divergent three-point functions, while grouping 
six-dimensional boxes and finite vertices into a single (finite) function.

If one wants to have a PO definition for the Higgs boson decaying into $\PZ\PGg$, one must
accept that the only completely consistent choice is $\PH \to \PZ_{\mathrm c}\PGg$, \ie the 
$\PZ$ at its complex pole, as discussed in \Brefs{Passarino:2010qk,Goria:2011wa}.
\section{Decay: numerics \label{Dec}}
We start by considering $\PH \to \Pep\Pem\PGg$ and introduce kinematical cuts as done in
\Bref{Dicus:2013ycd}: $M_{ij} > k_{ij}\,\mh$ with $i,j = \Pep,\Pem,\PGg$. Furthermore,
always following \Bref{Dicus:2013ycd}, we require that one fermion has energy greater than 
$25\UGeV$, the other greater than $7\UGeV$, while $E_{\PGg} > 5\UGeV$.
Note that in \Bref{Chatrchyan:2013vaa} a cut $M_{\Plp\Plm} > 50\UGeV$ is required.

A blind comparison (input parameters such as $\mt$ are not given in \Bref{Dicus:2013ycd})
gives a substantial agreement, at the level of few percentages.
Our results are given in \refT{tab:HTO_1}, for $\mh = 125\UGeV$.

The results of \refT{tab:HTO_1} should be compared with the SM (on-shell) prediction for 
$\Gamma(\PH \to \PZ\PGg)\times\mbox{Br}(\PZ \to \Pep\Pem)$, which is $0.214\UkeV$ and
with $\Gamma(\PH \to \PZ\PGg)= 9.27\UkeV$~\cite{Dittmaier:2011ti}. It may be of interest to
observe that $\Gamma\lpar \PH \to \Pep\Pem\Pep\Pem\rpar = 0.133\UkeV$.
\renewcommand{\arraystretch}{0.5}
\renewcommand{\tabcolsep}{15pt}
\begin{table}
\begin{center}
\caption[]{\label{tab:HTO_1}{Partial decay width for the process $\PH \to \Pep\Pem\PGg$ at
$\mh = 125\UGeV$, with cuts corresponding to $M_{ij} > k_{ij}\,\mh$ with $i,j = \Pep,\Pem,\PGg$.}}
\vspace{0.2cm}
\begin{tabular}{ccc}
\toprule
& & \\
$k$ & $\Gamma(\PH \to \Pep\Pem\PGg)/\Gamma(\PH \to \PGg\PGg) [\%]$ & $\Gamma[\UkeV]$ \\
& & \\
\midrule
& & \\
$0.1$ & $2.51$  & $0.2325$ \\
$0.2$ & $1.90$  & $0.1765$ \\
$0.3$ & $1.25$  & $0.1163$ \\
$0.4$ & $0.59$ & $0.0550$ \\
& & \\
$k_{\Pep\PGg} = k_{\Pem\PGg} = 0.1\; k_{\Pep\Pem} = 0.6$ & $2.03$ & $0.1884$ \\
& & \\
\bottomrule
\end{tabular}
\end{center}
\end{table}
\renewcommand{\tabcolsep}{6pt}
\renewcommand{\arraystretch}{1}
\paragraph{The process $\mathbf{\PH \to \PAf\Pf\PGg}$}
We have extended the calculation to include different fermions in the final state. First, we 
define cuts, following \Bref{Dicus:2013ycd}: 
\bqa
M_{\PAf\Pf} \equiv M\lpar \PAf\Pf\rpar > 0.1\,\mh \quad &
M_{\Pf\PGg} \equiv M\lpar \Pf\PGg\rpar > 0.1\,\mh       &
\quad 
M_{\PAf\PGg} \equiv M\lpar \PAf\PGg\rpar > 0.1\,\mh
\label{refcuts}
\eqa

With the cuts of \eqn{refcuts} we obtain the results shown in \refT{tab:HTO_2} for different
lepton and quark final states. It is worth noting that LO and NLO amplitudes do not interfere, 
as long as fermion masses are neglected in NLO, since the corresponding amplitudes belong to 
different helicity sets. For $\PGt$ and $\PQb$ the LO result is the leading one.
\renewcommand{\arraystretch}{0.5}
\renewcommand{\tabcolsep}{15pt}
\begin{table}[b]
\begin{center}
\caption[]{\label{tab:HTO_2}{Partial decay widths for the process $\PH \to \PAf\Pf\PGg$ at
$\mh = 125\UGeV$} and with cuts of \eqn{refcuts}. Both LO and NLO ($\mpf = 0$) results are 
shown.}
\vspace{0.2cm}
\begin{tabular}{lll}
\toprule
 &  & \\
$\Gamma_{\myLO} [\UkeV]$ & $\Gamma_{\myNLO} [\UkeV]$ & $\Pf$ \\
\midrule 
$0.29\,\times\,10^{-6}$ & $0.233$  & $\Pe$ \\
$0.012$                 & $0.233$  & $\PGm$ \\
$3.504$                 & $0.233$  & $\PGt$ \\
$0.013$                 & $0.874$  & $\PQd$ \\
$8.139$                 & $0.866$  & $\PQb$ \\
 &  & \\
\bottomrule
\end{tabular}
\end{center}
\end{table}
\renewcommand{\tabcolsep}{6pt}
\renewcommand{\arraystretch}{1}

A noteworthy effect of a finite $\mt$ can be seen in the $\PQb\,$-channel; for a $\PQb$
final state there are more Feynman diagrams contributing to the process due to the
fact that the $\PH$ boson has a non-zero coupling with top quarks.

The effect of a cut on $M_{\PAf\Pf}$, designed to enhance the contribution of the $\PZ$ 
peak, are given in \refT{tab:HTO_3}; here we fix the cuts such that $k_{\Pf\PGg} = k_{\PAf\PGg} = 
0.1$ and compare $k_{\PAf\Pf} = 0.1$ with $k_{\PAf\Pf} = 0.6$. The change corresponds to a $19\%$ 
reduction of the signal for the $\Pep\Pem\PGg$ final state.

Our calculation shows that $\Gamma(\PH \to \Pep\Pem\PGg)$, with $k_{ij} = 0.1$, is an increasing 
function of $\mh$. At $\mh = 120\UGeV$ we find 
$\Gamma(\PH \to \Pep\Pem\PGg)/\Gamma(\PH \to \PZ\PGg) = 3.9\%$
and $\Gamma(\PH \to \Pep\Pem\PGg)/\Gamma(\PH \to \PGg\PGg) = 2.0\%$ while at $\mh = 160\UGeV$
the ratios become $3.1\%$ and $4.5\%$ respectively.

\renewcommand{\arraystretch}{0.3}
\renewcommand{\tabcolsep}{15pt}
\begin{table}
\begin{center}
\caption[]{\label{tab:HTO_3}{The effect of kinematical cuts on $M_{\PAf\Pf}$ for the process 
$\PH \to \PAf\Pf\PGg$ at $\mh = 125\UGeV$}}
\vspace{0.2cm}
\begin{tabular}{lllll}
\toprule
 & & & & \\
$\Pf$ & $\Gamma_{\myNLO}[\UkeV]$ & & $\Gamma_{\myLO}[\UkeV]$ & \\
\midrule
 & & & & \\
 & $M_{\PAf\Pf} > 0.1\,\mh$ & $M_{\PAf\Pf} > 0.6\,\mh$ 
 & $M_{\PAf\Pf} > 0.1\,\mh$ & $M_{\PAf\Pf} > 0.6\,\mh$ \\
\midrule
$\PGm$ & $0.233$ & $0.188$ & $0.012$ & $0.010$ \\
$\PQd$ & $0.874$ & $0.835$ & $0.013$ & $0.011$ \\
$\PQb$ & $0.866$ & $0.831$ & $8.139$ & $6.745$ \\
 & & \\
\bottomrule
\end{tabular}
\end{center}
\end{table}
\renewcommand{\tabcolsep}{6pt}
\renewcommand{\arraystretch}{1}

\begin{figure}
\begin{minipage}{1.\textwidth}
  \vspace{-1.cm}
  \includegraphics[width=0.55\textwidth, bb = 0 0 595 842]{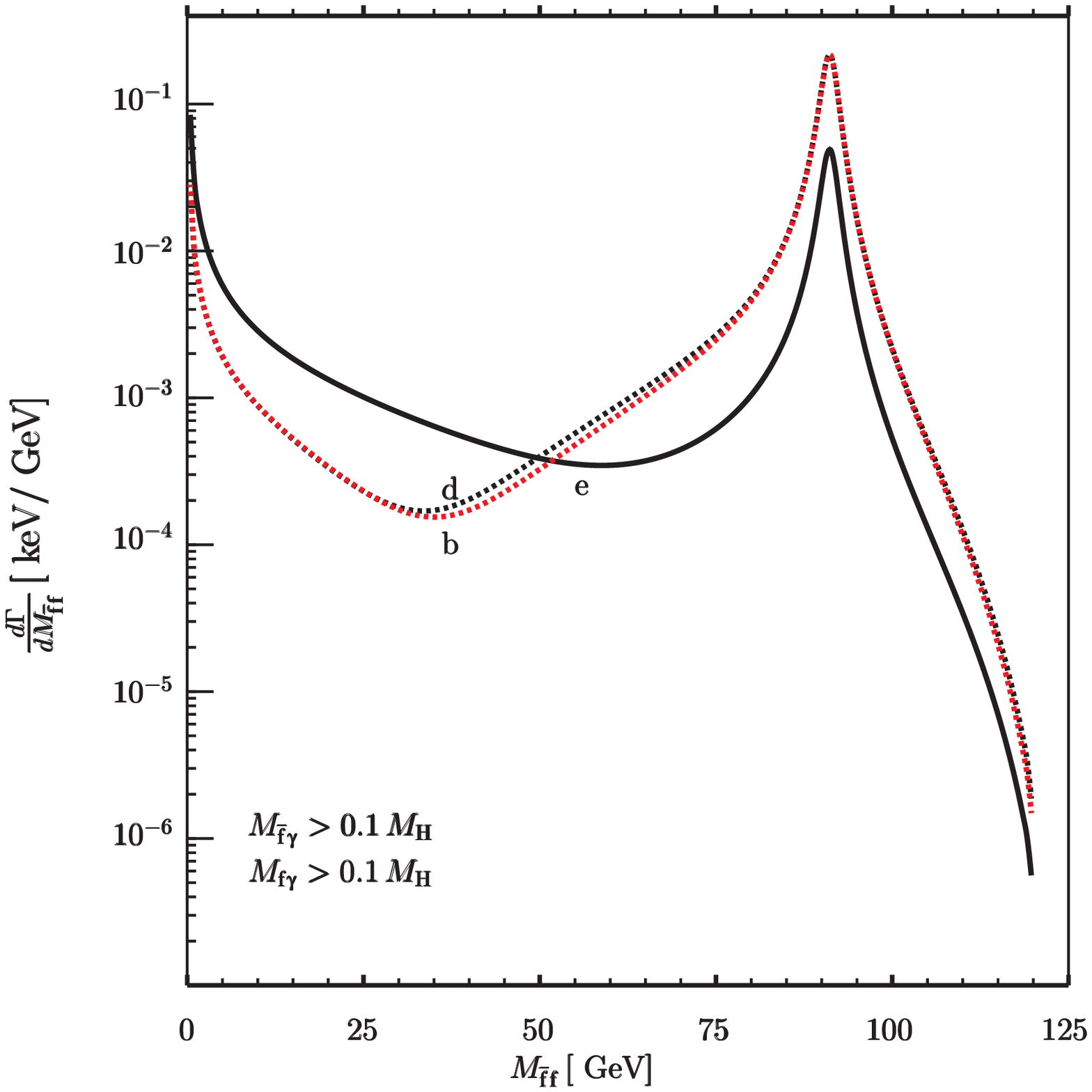}
  \hspace{-1.5cm}
  \includegraphics[width=0.55\linewidth, bb = 0 0 595 842]{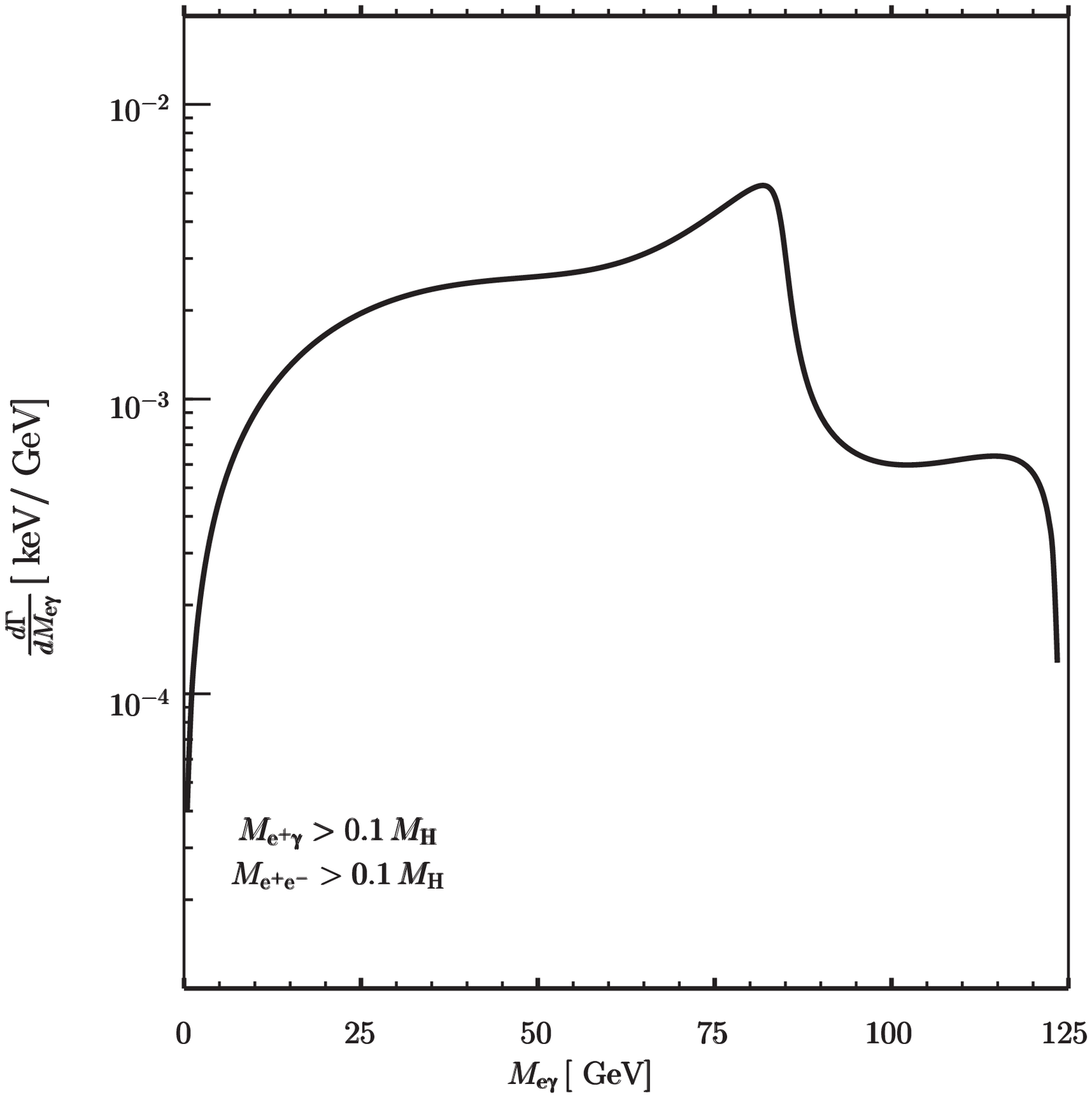}
  \vspace{-3.5cm}
  \caption[]{Invariant mass distributions in the process $\PH \to \PAf\Pf\PGg$ at
           $\mh = 125\UGeV$. The $M_{\PAf\Pf}$ distribution (left panel). 
           The $M_{\Pe\PGg}$ distribution (right panel). The case $\Pf = \PQb$ is shown for 
           comparison but the corresponding LO result is not included.} 
\label{fig:HTO_12}
\end{minipage}
\end{figure}

\begin{figure}[t]
\begin{minipage}{1.\textwidth}
  \vspace{-1.cm}
  \includegraphics[width=0.55\textwidth, bb = 0 0 595 842]{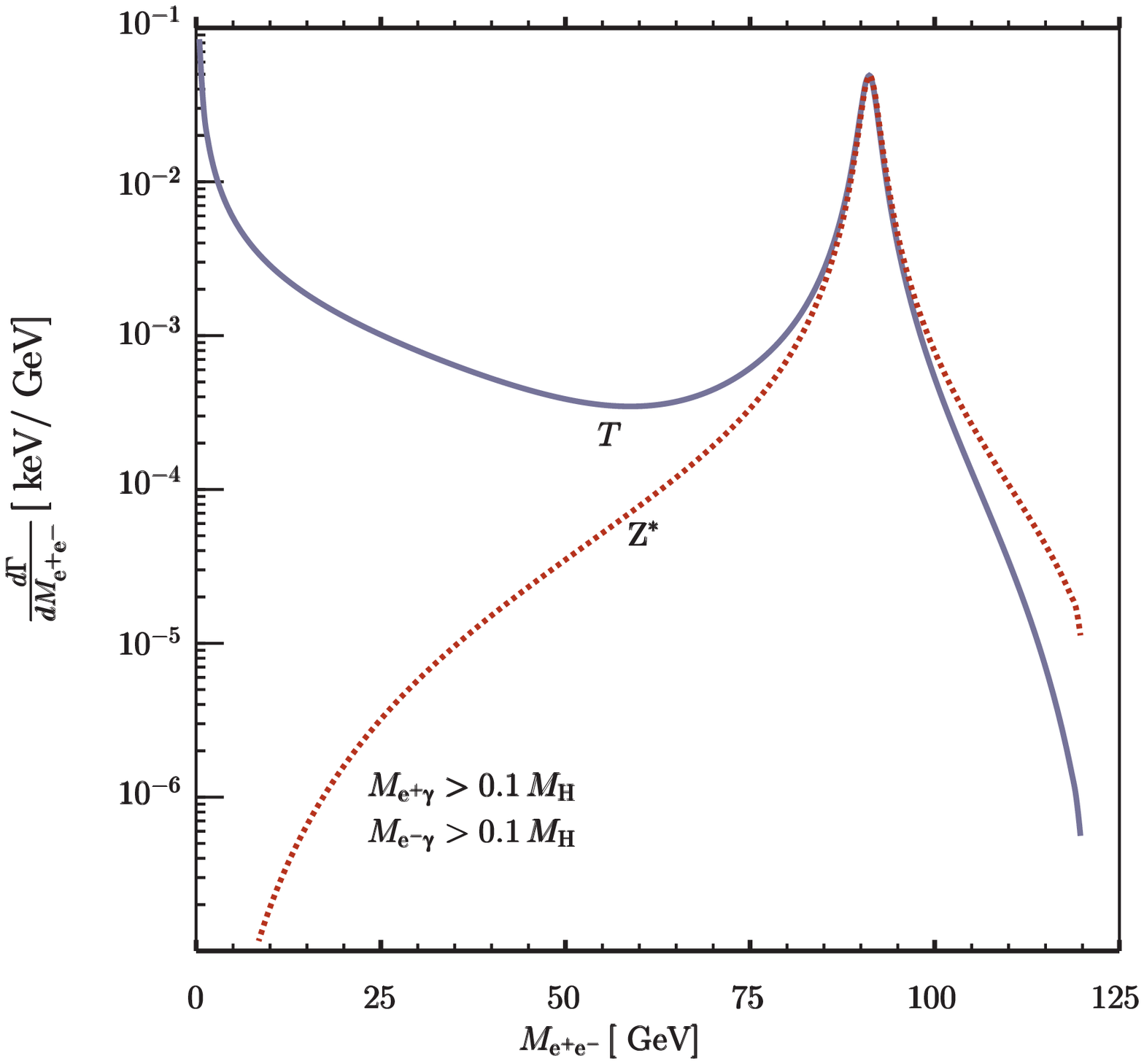}
  \hspace{-1.cm}
  \includegraphics[width=0.55\textwidth, bb = 0 0 595 842]{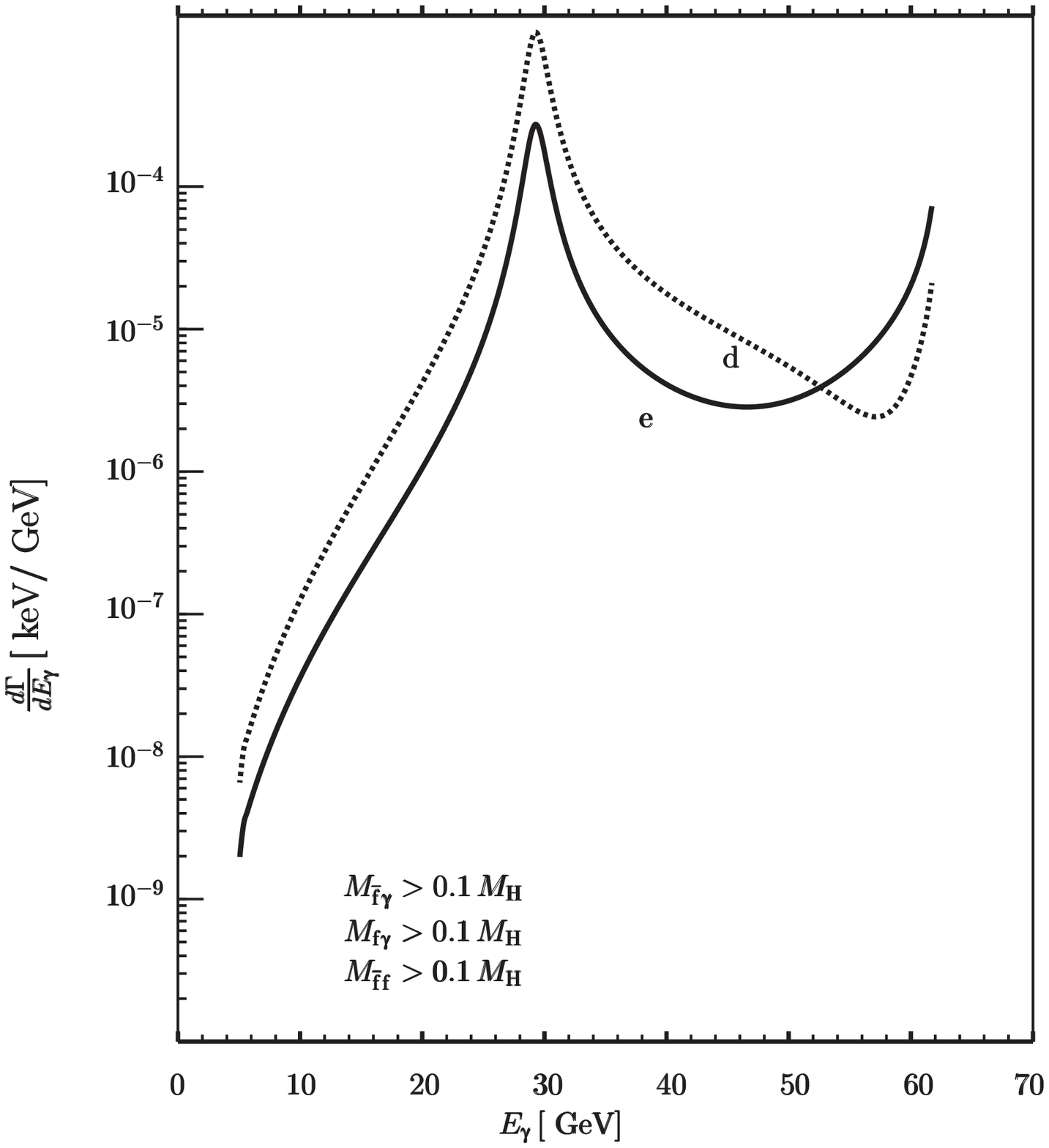}
  \vspace{-3.5cm}
  \caption[]{The process $\PH \to \PAf\Pf\PGg$ at $\mh = 125\UGeV$. 
           Comparing the total $M_{\PAf\Pf}$ distribution with the unphysical
           (off-shell) $\PZ^*$ component (left panel).
           The $E_{\PGg}$ distribution (right panel).} 

\label{fig:HTO_34}
\end{minipage}
\end{figure}

\begin{figure}[t]
\begin{minipage}{1.\textwidth}
  \vspace{-1.cm}
  \includegraphics[width=0.55\textwidth, bb = 0 0 595 842]{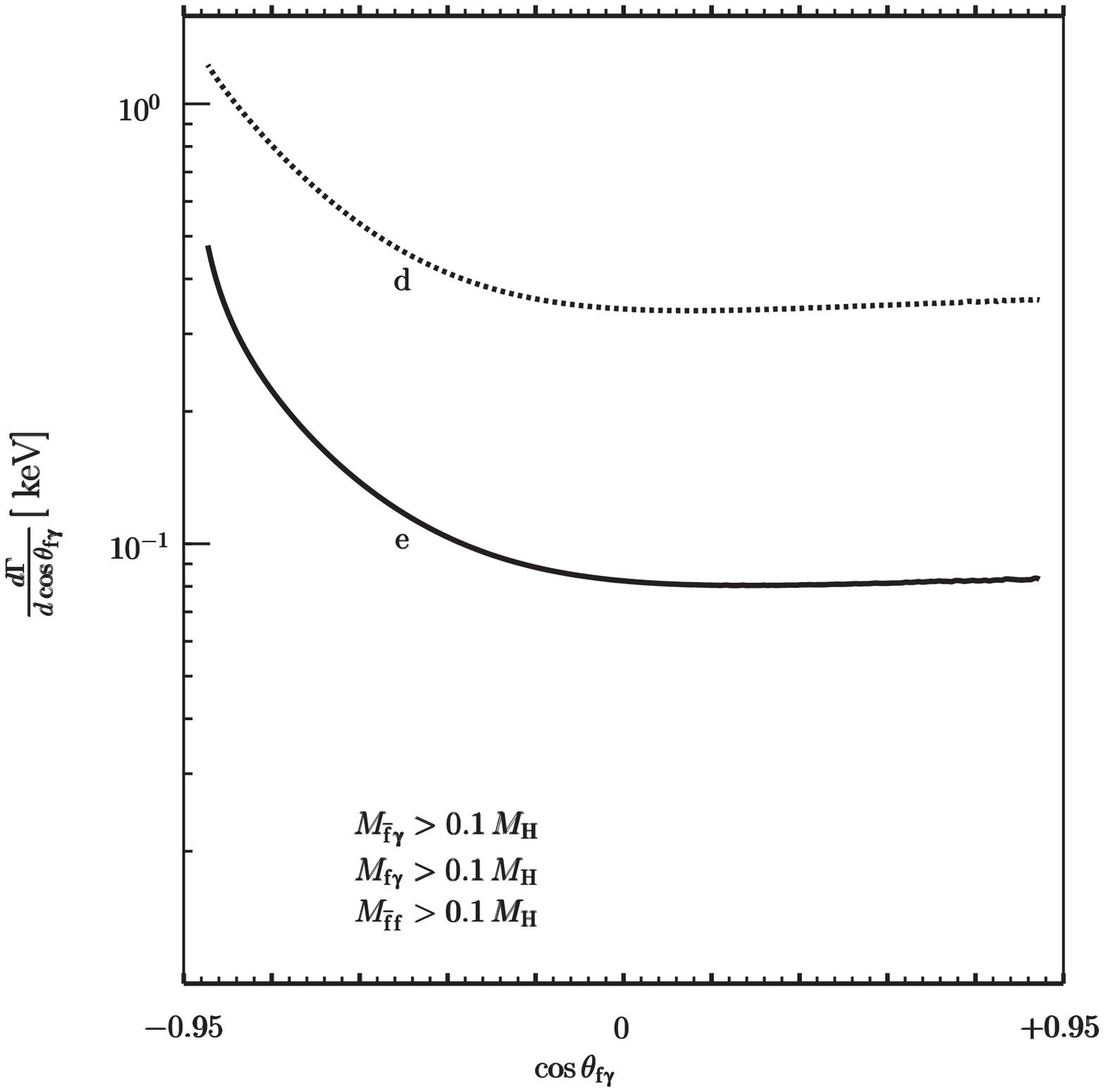}
  \hspace{-1.cm}
  \includegraphics[width=0.55\textwidth, bb = 0 0 595 842]{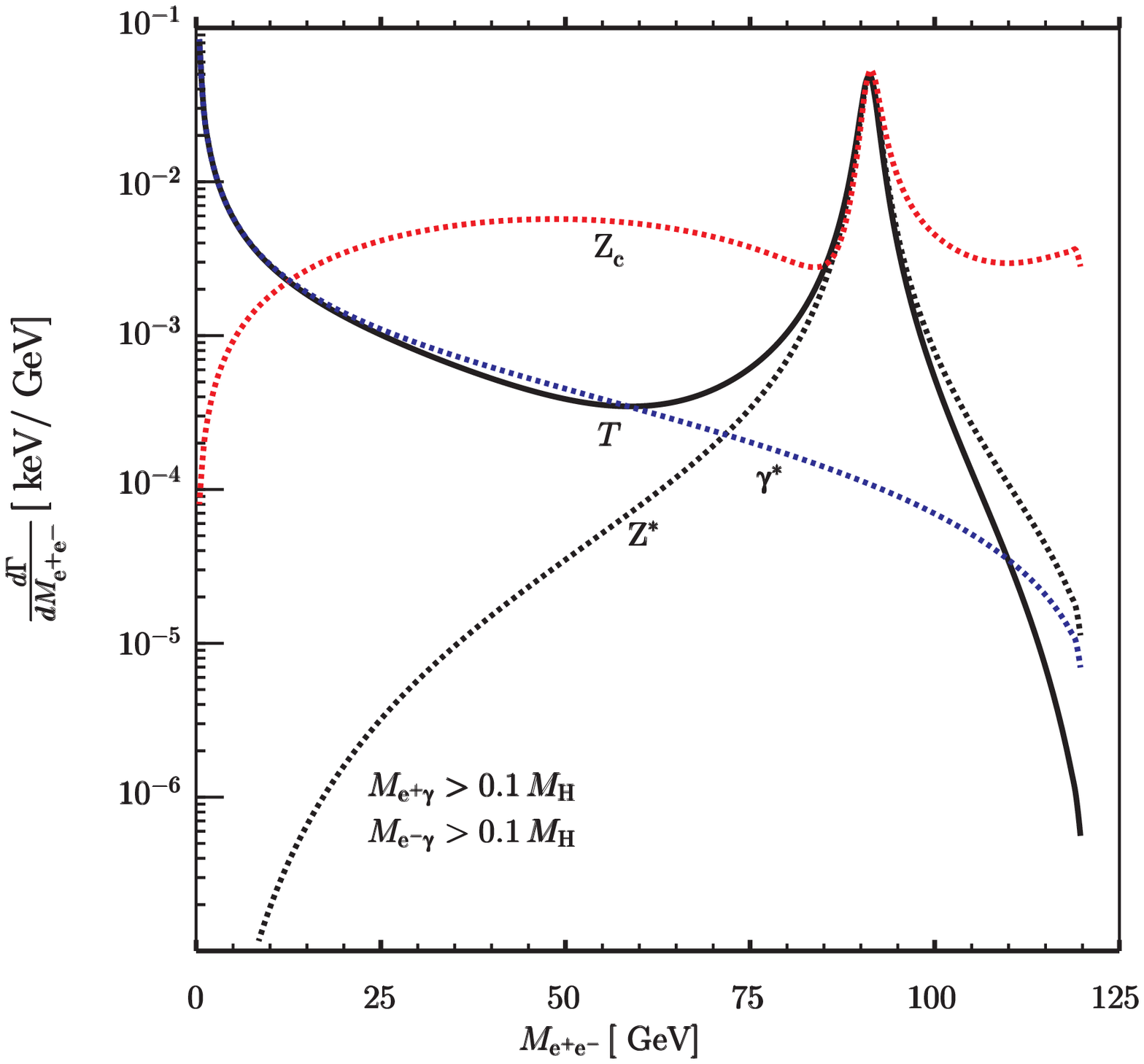}
  \vspace{-3.5cm}
  \caption[]{The $\cos\theta_{\Pf\PGg}$ distribution for $\PH \to \PAf\PF\PGg$ (left panel).
           Summary of the total $M_{\Pep\Pem}$ distribution in $\PH \to \Pep\Pem\PGg$ and of the
           partial components, $\PZ^*$, $\PGg^*$ and $\PZ_{\mathrm c}$ 
           (right panel).} 
\label{fig:HTO_56}
\end{minipage}
\end{figure}

We now study distributions; in Figures~\ref{fig:HTO_12}-\ref{fig:HTO_56} we show various 
distributions for the Dalitz decay of the Higgs boson at $125\UGeV$. The total $M_{\PAf\Pf}$ 
distribution is given in \refF{fig:HTO_12} (left panel), showing the $\PZ\,$-peak as well as 
the Coulomb peak at small values of the invariant $\PAf\Pf\,$-mass. The right panel of 
\refF{fig:HTO_12} gives the $M_{\Pe\PGg}$ distribution.

In \refF{fig:HTO_34} we compare the total $M_{\PAf\Pf}$ distribution with the unphysical
component of the decay, given by the off-shell $\PZ$ boson. Although the latter is a 
gauge-dependent quantity the figure gives a qualitative description of the $\PZ$ non-resonant 
background.

The $E_{\PGg}$ distribution for the $\PH \to \Pep\Pem\PGg$ decay is given in the right panel
of \refF{fig:HTO_34}, showing that the process is dominated by sufficiently hard photons,
with a maximum around $E_{\PGg} = 30\UGeV$.

In \refF{fig:HTO_56} (left panel) we show the angular distribution in terms of 
$\cos\theta_{\Pf\PGg}$; once again, the dominant contribution is given by non-collinear photons.
Finally, in \refF{fig:HTO_56}, we give a summary of the various (physical and unphysical) 
components in the decay $\PH \to \Pep\Pem\PGg$: the total ($T$), the off-shell ($\PZ^*$), the 
$\PGg^*$ and the $\PZ_{\mathrm c}$ ones.

\paragraph{Pseudo-observable}
As we have described above, the pseudo-observable of interest is $\Gamma_{\mathrm c} = 
\Gamma(\PH \to \PZ_{\mathrm c}\PGg \to \Pep\Pem\PGg)$ which we compare with $\Gamma_{\tot} =
\Gamma(\PH \to \Pep\Pem\PGg)$. Requiring $M_{\Pep\PGg} > 0.1\,\mh$ and
$M_{\Pem\PGg} > 0.1\,\mh$, we impose an additional cut on $M_{\Pep\Pem}$ around the $\PZ\,$-peak,
$\mz - \xi\,\tgz < M_{\Pep\Pem} < \mz + \xi\,\tgz$ and obtain the results 
shown in \refT{tab:HTO_4}.

\begin{figure}[t]
\begin{minipage}{1.\textwidth}
  \vspace{-1.cm}
  \includegraphics[width=0.55\textwidth, bb = 0 0 595 842]{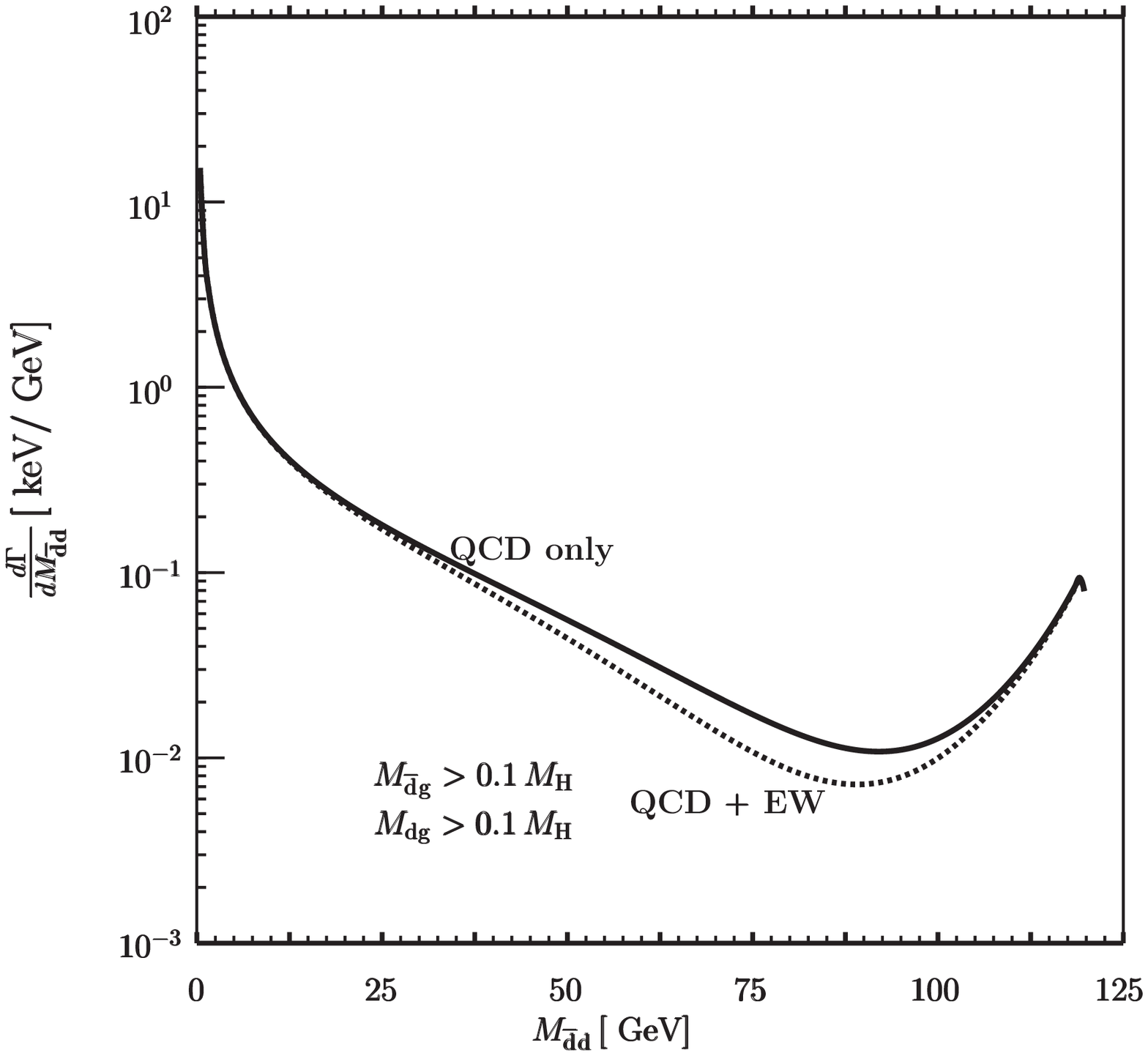}
  \hspace{-1.cm}
  \includegraphics[width=0.55\textwidth, bb = 0 0 595 842]{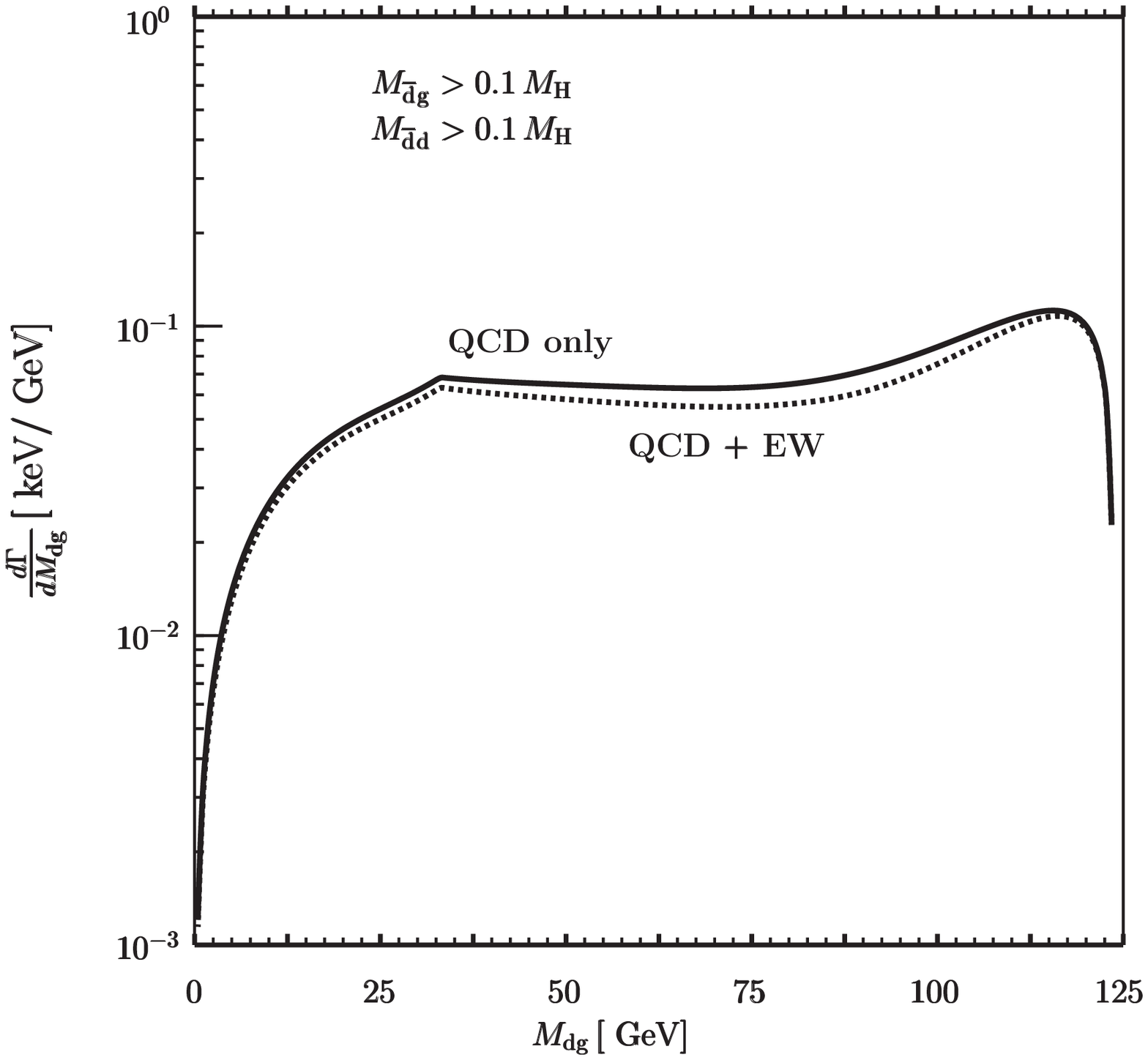}
  \vspace{-3.5cm}
  \caption[]{The process $\PH \to \PAQq\PQq\Pg$ at $\mh = 125\UGeV$.
           The $M_{\PAQd\PQd}$ distribution for the decay $\PH \to \PAQd\PQd\Pg$ (left panel).
           The $M_{\PQd\Pg}$ distribution for the decay $\PH \to \PAQd\PQd\Pg$ (right panel).}
\label{fig:HTO_78}
\end{minipage}
\end{figure}

\renewcommand{\arraystretch}{0.4}
\renewcommand{\tabcolsep}{15pt}
\begin{table}[b]
\begin{center}
\caption[]{\label{tab:HTO_4}{
The pseudo-observable $\Gamma_{\mathrm c} = \Gamma(\PH \to \PZ_{\mathrm c}\PGg \to \Pep\Pem\PGg)$ 
and $\Gamma_{\tot} = \Gamma(\PH \to \Pep\Pem\PGg)$. Here $M_{\Pep\PGg} > 0.1\,\mh$,
$M_{\Pem\PGg} > 0.1\,\mh$ and
$\mz - \xi\,\tgz < M_{\Pep\Pem} < \mz + \xi\,\tgz$.}}
\vspace{0.2cm}
\begin{tabular}{cccc}
\toprule
& & &\\
$\xi$ & $\Gamma_{\tot} [\UkeV]$ & $\Gamma_{\mathrm c} [\UkeV]$ &
$R_{\mathrm c} = \Gamma_{\mathrm c}/\Gamma_{\tot}$ \\
& & &\\
\midrule
& & &\\
$1$  & $138.7$ & $154.1$ & $1.11$ \\
$2$  & $166.2$ & $194.8$ & $1.17$ \\
$3$  & $176.4$ & $217.9$ & $1.24$ \\
$4$  & $181.7$ & $236.5$ & $1.30$ \\
$5$  & $185.0$ & $253.6$ & $1.37$ \\
& & & \\
\bottomrule
\end{tabular}
\end{center}
\end{table}
\renewcommand{\tabcolsep}{6pt}
\renewcommand{\arraystretch}{1}
The ratio $R_{\mathrm c}(\xi)$, defined in \refT{tab:HTO_4}, gives the correction factor
for extracting the pseudo-observable once a cut $\xi$ is selected around $M_{\PAf\Pf} =
\mz$.
\paragraph{The process $\mathbf{\PH \to \PAQq\PQq\Pg}$}
We have extended the calculation including a final state with a pair of light quarks and a gluon.
Each helicity amplitude contain a piece proportional to $g g^3_{\ssS}$ and a piece proportional
to $g^3 g_{\ssS}$ where $g$ is the $SU(2)$ coupling constant and $g_{\ssS}$ is the strong coupling
constant. Therefore, there are genuinely EW corrections to the result, not only the QCD triangle 
of top-quarks corresponding to $\PH \to \Pg^* (\to \PAQd\PQd) \Pg$. For a $\PQd\,$-quark with 
a cut $k_{ij} = 0.1$, $i,j = \PAQd,\PQd,\Pg$, we have a partial width (QCD + EW) of
$7.085\UkeV$ where the QCD part is $7.836\UkeV$ with an EW contribution of ${-}9.58\%$.
We present the $M_{\PAQd\PQd}, M_{\PQd\Pg}$ distributions in \refF{fig:HTO_78}. 
\section{Production: numerics \label{Pro}}
To discuss associated Higgs boson production at LHC ($8\UTeV$) we consider the following 
processes: $\PAQq + \PQq \to \PH + \Pg(\PGg)$, $\PQq(\PAQq) + \Pg \to \PH + \PQq(\PAQq)$;
here $\PQu$ stands for $\PQu \oplus \PQc$ and $\PQd$ for $\PQd \oplus \PQs$.

We study the total cross-section at $8\UTeV$ as well as the $\pT$ distribution of the
parton in the final state (gluon or quark). All processes are computed at NLO accuracy, which is
the leading contribution for massless light quarks. Renormalization and factorization QCD scales 
are fixed at $\muR = \muF = \mh$ and their variation is postponed until \refS{THU}.
\paragraph{Cross sections}
In \refT{tab:HTO_5} we show all the cross sections at $\sqrt{s} = 8\UTeV$, $\mh = 125\UGeV$, 
with a cut of $30\UGeV < \pT < 300\UGeV$, comparing QCD with QCD+EW (no cut on pseudo-rapidity 
applied); we can split the amplitude into a part proportional to $g g^3_{\ssS}$ and a part 
proportional to $g^3 g_{\ssS}$: the former is what we define as (hard) QCD component of the 
$\mpf = 0$ NLO amplitudes. The effect of NLO EW corrections is parametrized in terms of the 
relative deviation, $\delta_{\myEW} = \sigma_{\myQCD+\myEW}/\sigma_{\myQCD} - 1$.

It is worth noting that the $\PAQq\PQq\,$-annihilation cross sections are tiny (also due to
parton luminosity) while the $\PQq\Pg\,$-annihilation is enhanced, also by the contribution
of the gluon exchange in the $t\,$-channel (the vertex diagram).
The effect of including the EW part is larger in the annihilation channel where, however, the 
cross sections are much smaller, $38.8\Ufb$ for light quarks as compared to $2.4\Upb$ for the 
quark-gluon channel. For the latter there is a partial cancellation between $\PQq\Pg$ and
$\PAQq\Pg$.

Note that we do not discuss the case $\PQq = \PQb$ at 
LO + NLO QCD~\cite{Dawson:2004sh,Dicus:1998hs,Maltoni:2003pn,Boos:2003yi,Harlander:2003ai},
corresponding to a non-zero value of $\mb$ (see \Bref{Brein:2010xj} for more details). 
This part of the NLO corrections contains the soft/collinear QCD that can be added
incoherently to our result.

\begin{figure}[t]
\begin{minipage}{1.\textwidth}
  \includegraphics[width=0.55\textwidth, bb = 0 0 595 842]{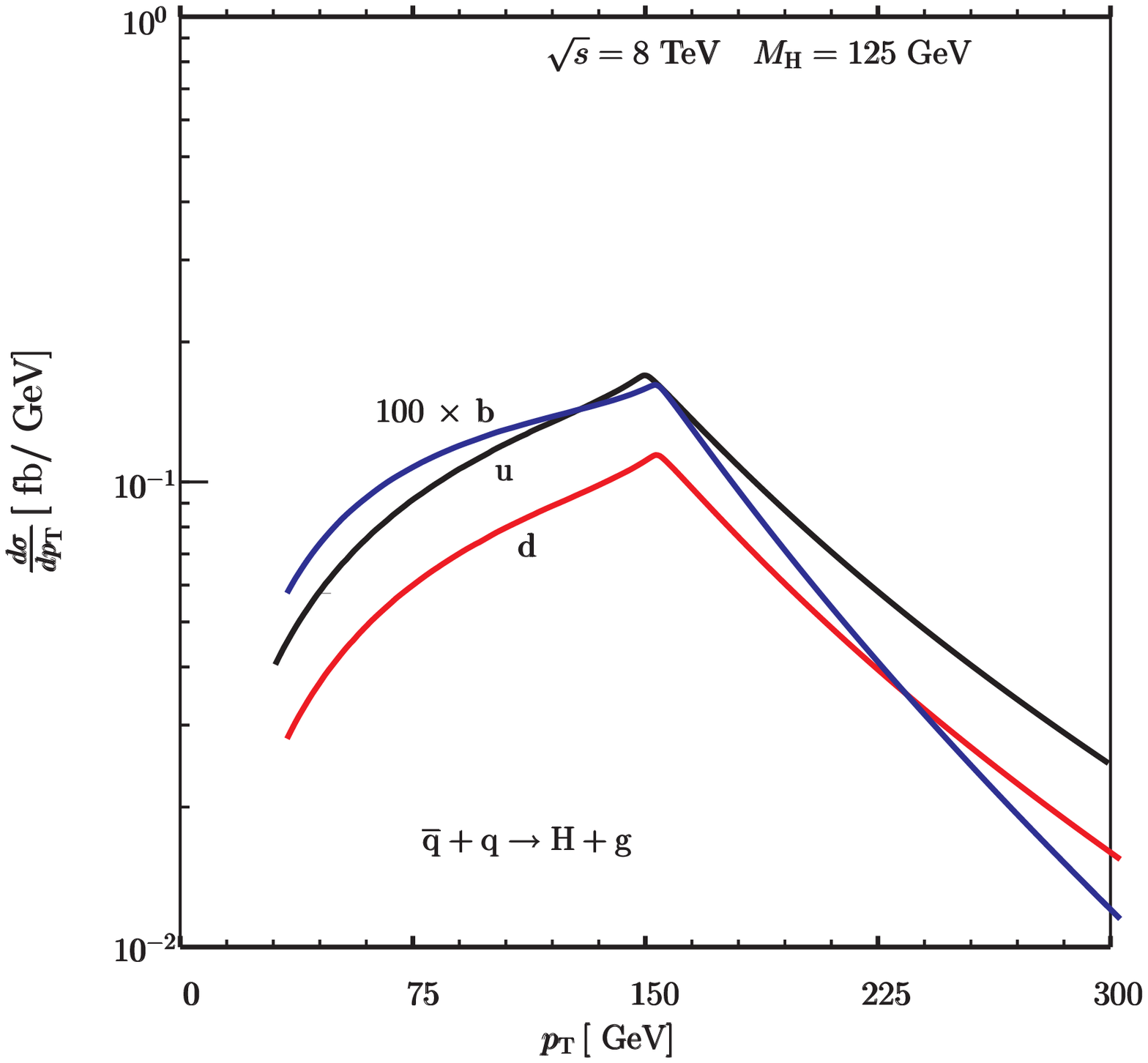}
  \hspace{-1.cm}
  \includegraphics[width=0.55\textwidth, bb = 0 0 595 842]{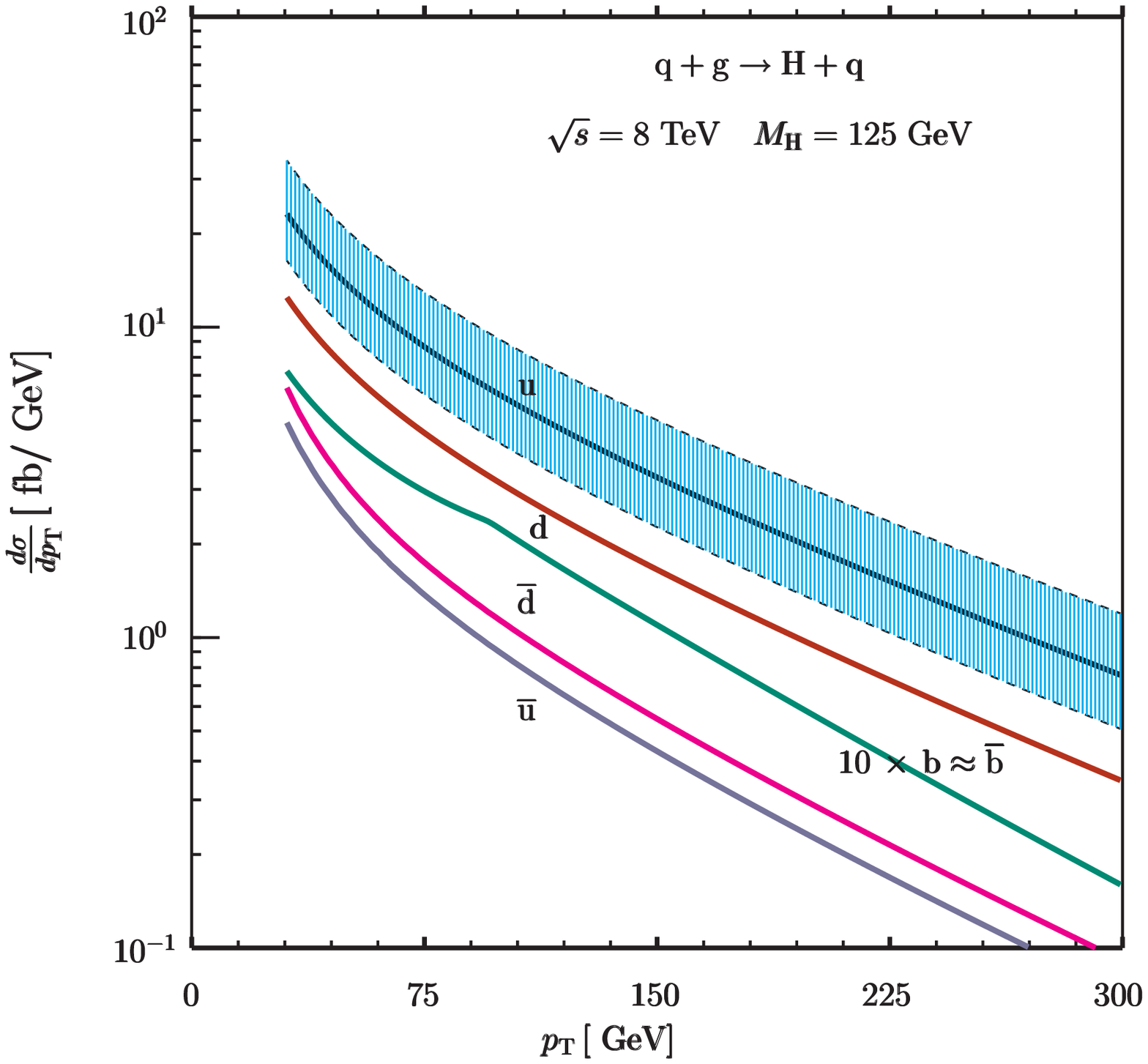}
  \vspace{-4.cm}
  \caption[]{The $\pT\,$-distribution for $\PAQq + \PQq \to \PH + \Pg$ at $\sqrt{s} = 8\UTeV$
   (left panel). 
   The $\pT\,$-distribution for $\PQq + \Pg \to \PH + \PQq$ at $\sqrt{s} = 8\UTeV$
   (right panel). The $\PQb\,$-quark is added for comparison but the corresponding LO + NLO 
   soft/virtual is not included. Visible is the effect of the $\mw+\mt$ normal threshold.}
\label{fig:HTO_PT12}
\end{minipage}
\end{figure}

If a cut on pseudo-rapidity, $\mid\eta\mid < 2.5$, is applied we register a reduction of 
$\approx 40\%$ on the cross sections. 
\renewcommand{\arraystretch}{0.3}
\renewcommand{\tabcolsep}{15pt}
\begin{table}[h]
\begin{center}
\caption[]{\label{tab:HTO_5}{Total cross sections for associate Higgs production at 
$\sqrt{s} = 8\UTeV$ and $\mh = 125\UGeV$ for $30\UGeV < \pT < 300\UGeV$.}}
\vspace{0.2cm}
\begin{tabular}{crrr}
\toprule
& & \\
\mbox{process} & $\sigma_{\myQCD+\myEW} [\Ufb]$ & $\sigma_{\myQCD} [\Ufb]$ & 
                 $\delta_{\myEW} [\%]$\\
& & \\
\midrule
& & \\
$\PAQu + \PQu \to \PH + \Pg$  & $23.24$  &  $26.25$ & ${-}11.5$ \\
$\PAQd + \PQd \to \PH + \Pg$  & $15.54$  &  $17.71$ & ${-}12.3$ \\
& & \\
Total                         & $38.78$  &  $43.96$ & ${-}11.8$ \\
& & \\
\midrule
$\PAQb + \PQb \to \PH + \Pg$  & $0.221$  &  $0.317$ & ${-}30.3$ \\
& & \\
\midrule
& & \\
$\PQu + \Pg \to \PH + \PQu$   & $1284.5$ &  $1312.3$ & ${-}2.1$  \\
$\PAQu + \Pg \to \PH + \PAQu$ & $203.2$  &  $192.1$  & ${+}5.8$ \\
$\PQd + \Pg \to \PH + \PQd$   & $668.0$  &  $684.7$  & ${-}2.4$  \\
$\PAQd + \Pg \to \PH + \PAQd$ & $259.5$  &  $242.5$  & ${+}7.0$ \\
& & \\
Total                         & $2415.2$ &  $2431.6$ & ${-}0.07$  \\ 
& & \\
\midrule
$\PQb + \Pg \to \PH + \PQb$   & $41.81$  &  $33.78$  & ${+}23.8$ \\
$\PAQb + \Pg \to \PH + \PAQb$ & $42.89$  &  $33.78$  & ${+}27.0$ \\
& & \\
\bottomrule
\end{tabular}
\end{center}
\end{table}
\renewcommand{\tabcolsep}{6pt}
\renewcommand{\arraystretch}{1}
\paragraph{$\mathbf{\pT}\,$-distributions}
We have analyzed the $\pT\,$-distribution for different processes.
In \refF{fig:HTO_PT12} (left panel) we show the $\pT\,$-distribution for 
$\PAQq + \PQq \to \PH + \Pg$ at $\sqrt{s} = 8\UTeV$ and $\mh = 125\UGeV$. 

In \refF{fig:HTO_PT12} (right panel) we show the $\pT\,$-distribution for 
$\PQq(\PAQq) + \Pg \to \PH + \PQq(\PAQq)$ at $\sqrt{s} = 8\UTeV$ and $\mh = 125\UGeV$. 
To illustrate the effect of QCD scales we have included the band corresponding to
$\PQu + \Pg \to \PH + \PQu$ for $\muR=\muF \in [\mh/2\,,\,2\,\mh]$.

Some of the features of the $\pT\,$-distributions can be understood by introducing
$\pT^2= \rho\,s$, the partonic variable $\shat= z\,s$ and the scaled Higgs mass $\mh =
\mu_{\PH}\,\sqrt{s}$. For $\rho$ fixed ($0 \le \rho \le (1-\mu^2_{\PH})^2/4$) we have
$z_+ \le z \le 1$ where $z_+ = \mu^2_{\PH} + 2\,\rho + 2\,\sqrt{\rho\,(\rho + \mu^2_{\PH})}$.
Cuts of the amplitudes are at $\shat = 4\,M^2_{\PQt}, 4\,M^2_{\PZ}$ \etc
The value $\shat = 4\,M^2_{\PQt}$ corresponds to $\pT= 149.86\UGeV$, reflecting the spike in the
$\pT\,$-distribution (crossing a normal threshold in $\shat$).
Similarly, the spike in \refF{fig:HTO_PT12} for $\PQb + \Pg \to \PH + \PQb$ corresponds to
the threshold $\shat = (\mw + \mt)^2$.

\begin{figure}
\begin{minipage}{1.\textwidth}
  \includegraphics[width=0.55\textwidth, bb = 0 0 595 842]{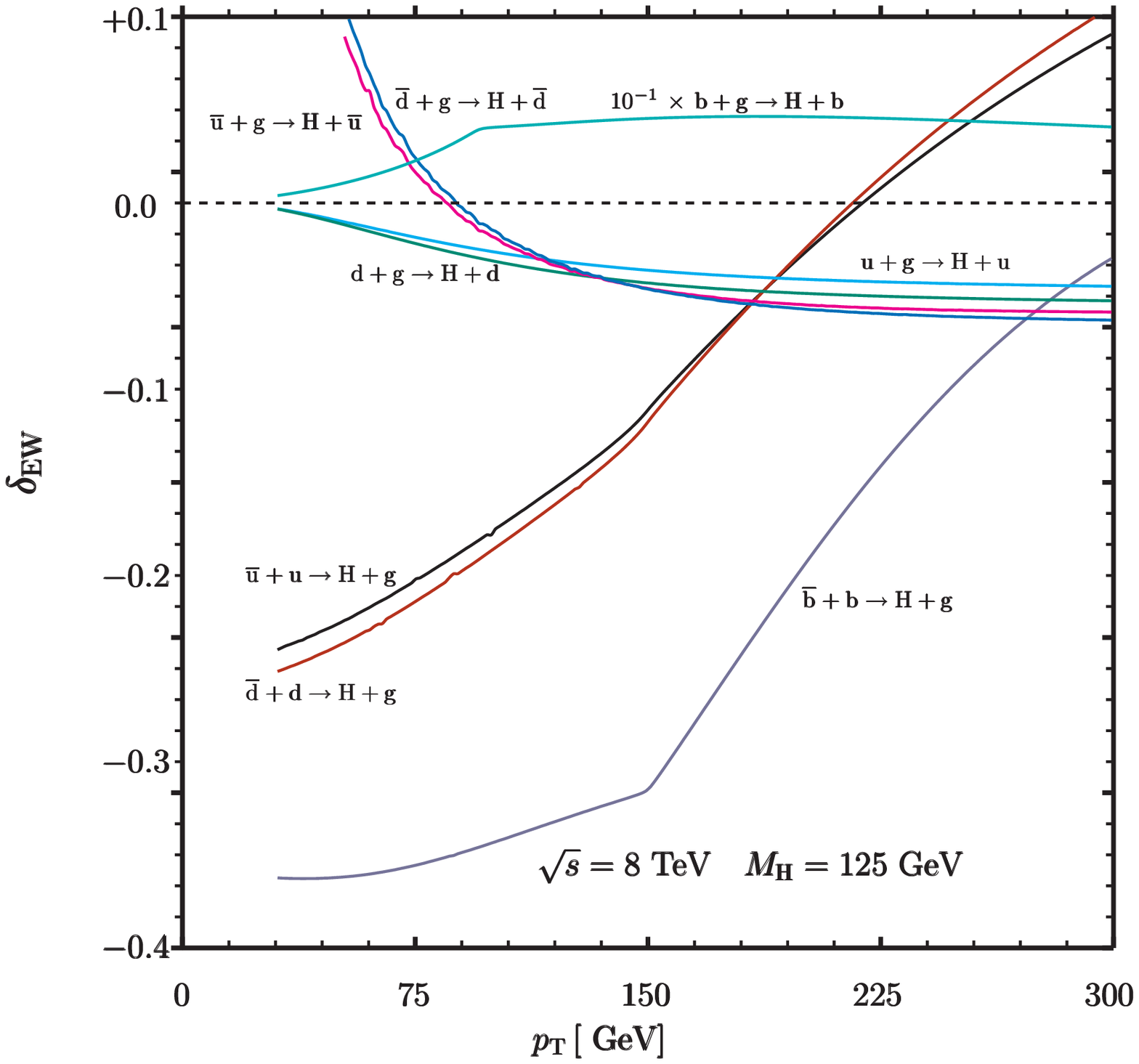}
  \hspace{-1.cm}
  \includegraphics[width=0.55\textwidth, bb = 0 0 595 842]{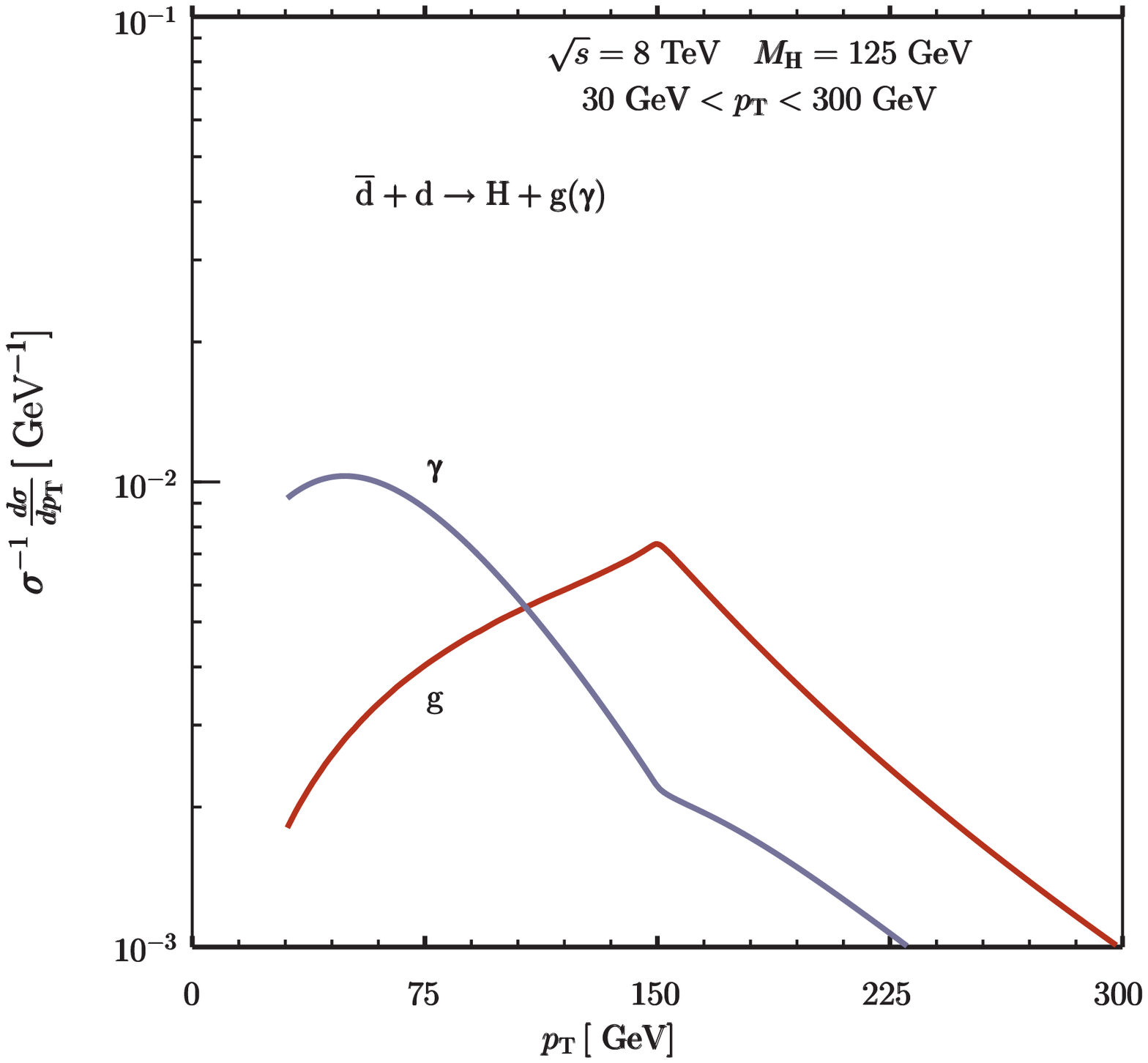}
  \vspace{-4.cm}
  \caption[]{EW effect, parametrized by $\delta_{\myEW} = 
           \sigma_{\myQCD+\myEW}/\sigma_{\myQCD} - 1$ (left panel). 
           Normalized $\pT\,$-distributions ($30\UGeV < \pT < 300\UGeV$) for 
           $\PAQd + \PQd \to \PH + \Pg$ and $\PAQd + \PQd \to \PH + \PGg$.}
\label{fig:HTO_per_AG}
\end{minipage}
\end{figure}

The percentage effects of the EW component are summarized in the left panel of 
\refF{fig:HTO_per_AG} where we show $\delta_{\myEW}(\pT)$ for the $\PAQq\PQq\,$-channel and 
for the $\PQq(\PAQq)\Pg\,$-channel. 
For the annihilation channel the effect of including EW components reaches $-25\%$ for
$\pT$ around $30\UGeV$; note that $\delta_{\myEW}$ becomes positive around $\pT = 225\UGeV$ with
a $+10\%$ at $\pT = 300\UGeV$.
For the process $\PQq(\PAQq) + \Pg \to \PH + \PQq(\PAQq)$ we note the different behavior in 
the two channels $\PQq\Pg$ and $\PAQq\Pg$ and the large EW effects in the $\PQb\Pg\,$-channel
where, once again, we have not included LO and soft/collinear NLO.
\paragraph{The process $\PAQq + \PQq \to \PH + \PGg$}
This process is highly suppressed, being of purely EW origin; for $\PAQd + \PQd \to \PH + \PGg$
and $30\UGeV < \pT < 300\UGeV$ we find $\sigma= 0.052\Ufb$. The consequences of
a central photon requirement make the channel $\Pp\Pp \to \PH \PGg$ a process worthwhile
to investigate, although the smallness of the signal makes it questionable to discern
signal from background in $\PAQq + \PQq \to \PAQb + \PQb + \PGg$ or $\Pg + \Pg \to
\PAQb + \PQb + \PGg$~\cite{Abbasabadi:1997zr}. 

Higgs boson production in association with a photon via weak boson fusion has received
considerable attention in the literature, see \Bref{Arnold:2010dx}.
In \Bref{Gabrielli:2007zp} this process has been proposed to probe the $\PQb\,$-quark parton 
densities.

In the right panel of \refF{fig:HTO_per_AG} we compare the normalized $\pT\,$-distributions 
($30\UGeV < \pT < 300\UGeV$) for $\PAQd + \PQd \to \PH + \Pg$ and $\PAQd + \PQd \to \PH + \PGg$, 
showing that the $\PGg\,$-spectrum is softer than the $\Pg$ one. In both cases we have a spike
corresponding to the $\shat = 4\,M^2_{\PQt}$ normal threshold.
\paragraph{Production and decay}
Finally, we consider the full (signal) process, $\Pp\Pp \to \Pg\Pg \to \PH \to \Pep\Pem\PGg$,
going beyond the zero-width approximation. We can distinguish between a DZWA,
$\sigma\lpar \Pg\Pg \to \PH\rpar\,\times\,
\mathrm{BR}\lpar \PH \to \PZ\PGg\rpar\,\times\,
\mathrm{Br}\lpar \PZ \to \Pep\Pem\rpar$ and a ZWA
$\sigma\lpar \Pg\Pg \to \PH\rpar\,\times\,
\mathrm{BR}\lpar \PH \to \Pep\Pem\PGg\rpar$.

\begin{figure}
\begin{minipage}{1.\textwidth}
  \includegraphics[width=0.9\textwidth, bb = 0 0 595 842]{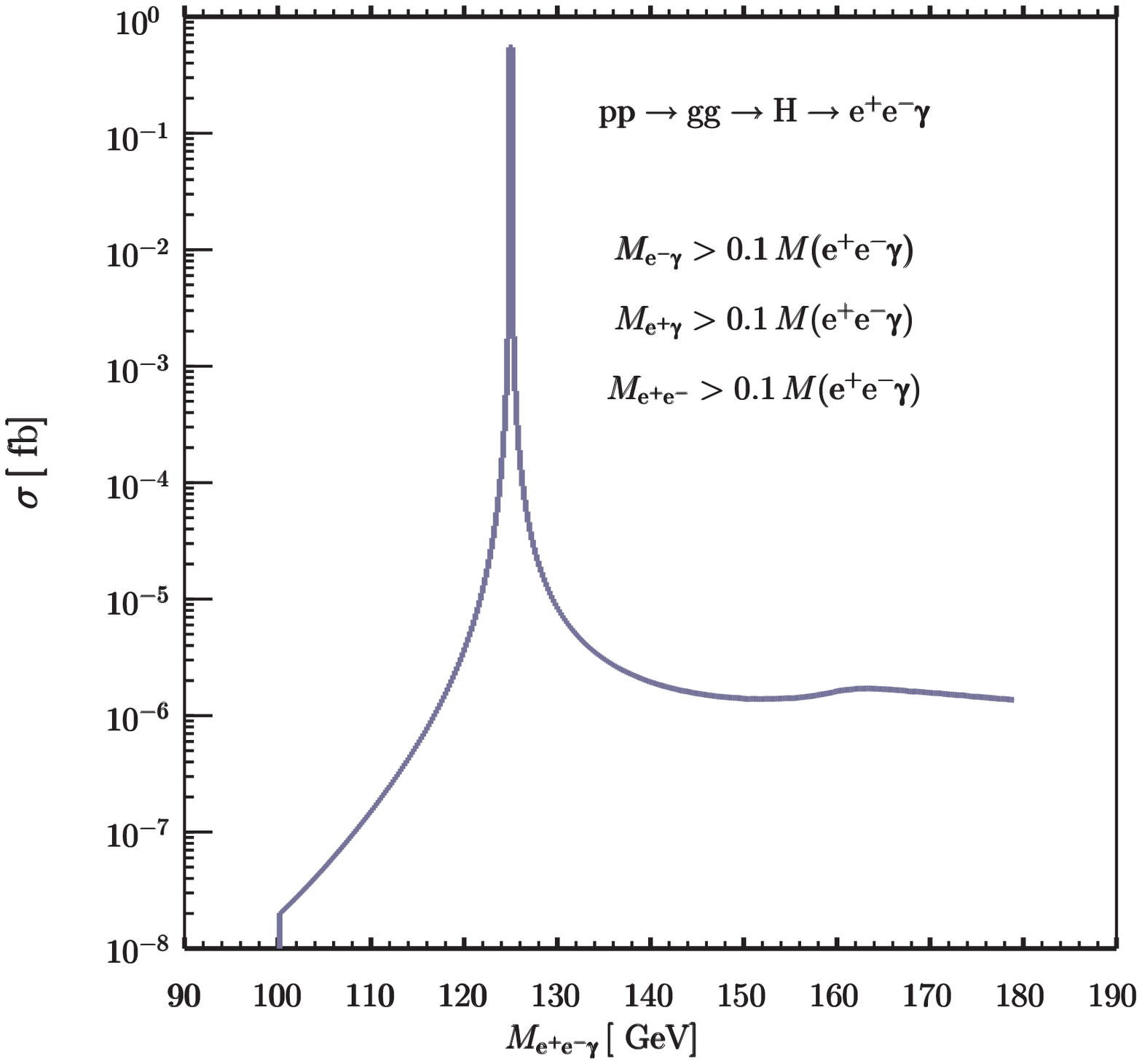}
  \vspace{-3.5cm}
  \caption[]{The invariant mass distribution for
             $\Pp\Pp \to \Pg\Pg \to \PH \to \Pep\Pem\PGg$.
             All finals state invariant masses are larger than $0.1\,M_{\Pep\Pem\PGg}$
             and a bin size of $200\UMeV$ is used.}
\label{fig:HTO_cat}
\end{minipage}
\end{figure}

We require that all finals state invariant masses are larger than $0.1\,M_{\Pep\Pem\PGg}$
and a bin size of $200\UMeV$ is used; the result is shown in \refF{fig:HTO_cat}. The calculation
is performed within the CPS-scheme~\cite{Actis:2006rc,Actis:2008uh,Passarino:2010qk} as
implemented in \Bref{Goria:2011wa}.
The distribution is asymmetric with a large tail for values of the invariant mass above $\mh$.
This is a known effect, described for the first time in \Bref{Kauer:2012hd}; it has to do with
the growth of the partial decay width with growing invariant masses, extending well above the
spike due to the $\PW\PW$ normal threshold. One has $\Gamma\lpar \PH \to \Pep\Pem\PGg\rpar) =
0.022\UkeV$ at $\mh = 100 \UGeV$ and $\Gamma\lpar \PH \to \Pep\Pem\PGg\rpar) = 6.91\UkeV$ at
$\mh = 190\UGeV$.
Of course, we are not claiming observability of the tail ($\approx 10^{-6}\Ufb$).
The main point of this exercise is to compare the full-fledged cross section at $8\UTeV$ with 
the corresponding DZWA and ZWA, although experimentally it will be very hard to construct an 
hypothesis test which can resolve ZWA versus the shape of \refF{fig:HTO_cat}, especially in 
this channel. A measure of the effect described in \Bref{Kauer:2012hd} looks more promising
in $\PH \to 4\,\Pl$ where the background is orders of magnitude lower.
\[
\begin{array}{ccr}
\cmidrule{1-3}
100\UGeV < M_{\Pep\Pem\PGg} < 180\UGeV           & 
M_{ij} > 0.1\,M_{\Pep\Pem\PGg}                   &
\sigma = 1.13\Ufb                                \\
\sigma\lpar \Pg\Pg \to \PH\rpar = 19.49\Upb              &
\mathrm{BR}\lpar \PH \to \PZ\PGg\rpar\,\times\,
\mathrm{Br}\lpar \PZ \to \Pep\Pem\rpar = 5.24\,10^{-5}   &
\sigma_{\mbox{\scriptsize{DZWA}}} = 1.02\Ufb              \\
\cmidrule{1-3}
\end{array}
\]

\section{Theoretical uncertainties \label{THU}}
For the decay $\PH \to \PAl\Pl\PGg$ the parametric uncertainties (PU) are tiny and we do not expect
any source of enhancement from missing higher orders (MHO), as long as one stays in the light 
Higgs region. Therefore, we expect MHOs of the same size as in $\PH \to \PGg\PGg$, \ie
$\le \ord{5\%}$.

For the decay $\PH \to \PAQq\PQq\Pg$ there is a ${\pm} 3.3\%$ effect when varying $\alphas$ by
${\pm} 0.0014$. We expect the effects of MHO to be the of the same order of those in
$\PH \to \Pg\Pg$, see \Brefs{Dittmaier:2011ti,Dittmaier:2012vm}.

For associated production the estimate is much less precise and usually not discussed in the
literature: consider the sub-process with the largest cross section, $\PQu + \Pg \to \Ph + \PQu$, 
the usual strategy of varying the QCD scales gives large effects, as shown in the right panel 
of \refF{fig:HTO_per_AG} where we adopt the standard recipe $\muR=\muF \in [\mh/2\,,\,2\,\mh]$.

This is expected since NNLO corrections are missing and NNLO is the first level of precision 
where one should start discussing uncertainties. According to the point of view expressed in 
\Bref{David:2013gaa} we are not going to use QCD scale variation as the true estimator of 
theoretical uncertainty; however, the MHO uncertainty clearly contains QCD scale variation and 
we must conclude that, at the present level of knowledge, these processes suffer from a large 
uncertainty, both in the total cross section and in the $\pT\,$-distribution.
\section{Conclusions}
In this work we provide a general framework for studying production and decay mechanisms of
the SM Higgs boson which are Yukawa suppressed at LO but not at NLO. The three-body decay
of the Higgs boson, $\PH \to \PAf\Pf\PGg(\Pg)$, is naturally framing the extraction of
pseudo-observables (\eg $\PH \PZ\PGg$) that have universal inherent meaning and are used 
in extracting information for the couplings of the newly discovered resonance.
Some of these effects have been studied separately in 
\Brefs{Abbasabadi:1996ze,Abbasabadi:2006dd,Abbasabadi:2004wq,Dicus:2013ycd} and
in \Brefs{Keung:2009bs,Brein:2010xj}; we have completed all calculations (updating the 
computational framework), extending previous results to give comprehensive view of the 
implications, including a comparison of the cross section for $\Pp\Pp \to \Plp\Plm\PGg$ at 
$8\UTeV$ with the corresponding zero-width approximation.

\section{Acknowledgments}
Significant discussions with S.~Actis, A.~David, M.~Duhrssen, A.~Maier, C.~Mariotti, D.~Rebuzzi,
M.~Spira and R.~Tanaka are gratefully acknowledged.
\clearpage
\bibliographystyle{atlasnote}
\bibliography{Dalf}{}

\end{document}